\documentclass[journal]{IEEEtran}
\usepackage{amsmath}
\usepackage[pdftex]{graphicx}
\usepackage{subfigure}
\usepackage{amsmath}
\usepackage{amssymb}
\usepackage{amsfonts}
\usepackage{lmodern}
\usepackage{float}
\usepackage{bm}
\usepackage{comment}
\usepackage{color}
\usepackage[pdftex]{hyperref}

\newcommand\orcidAuthor[1]{
\hspace*{-1mm}\includegraphics[keepaspectratio,width=0.7em]{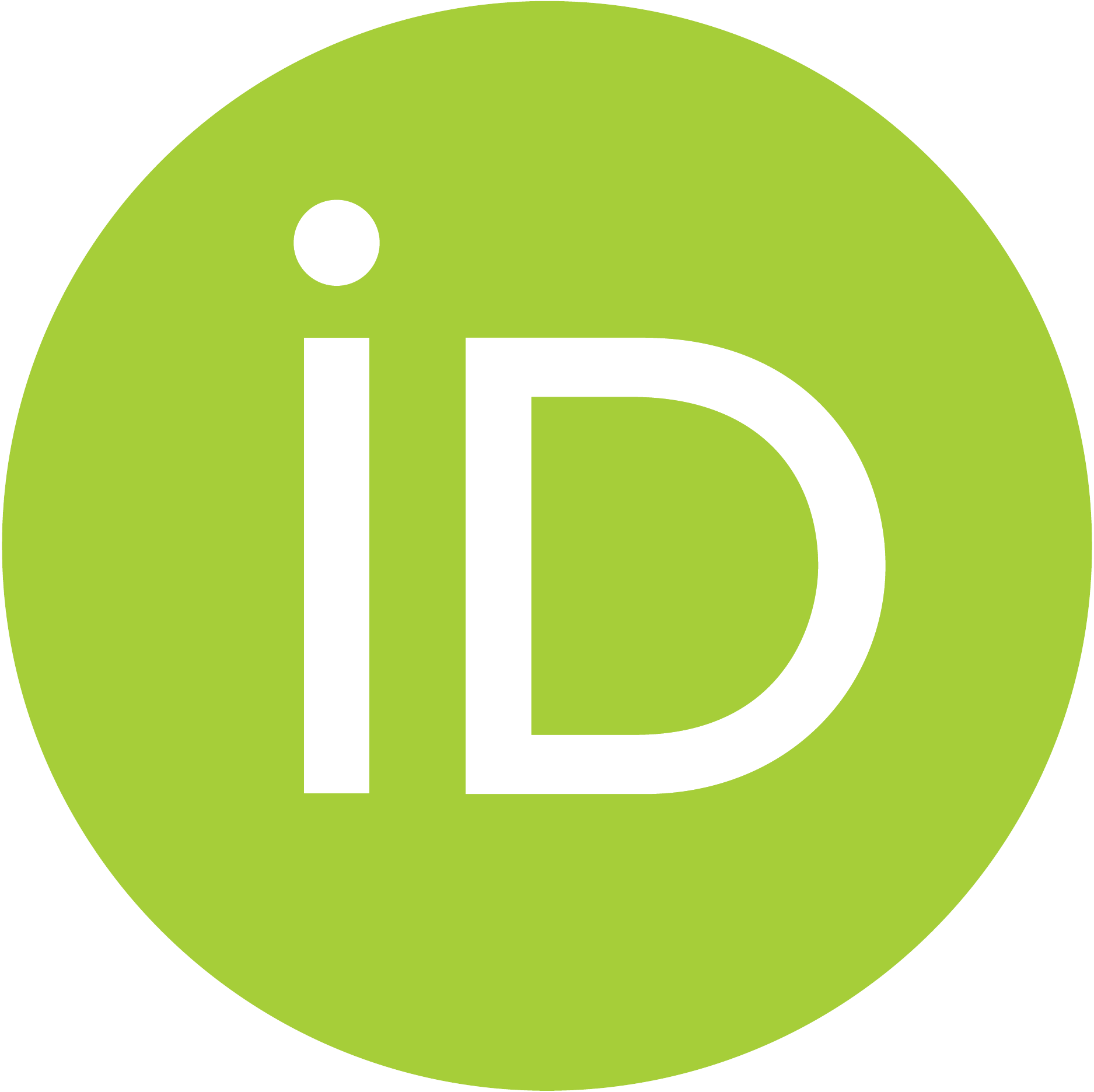}\href{https://orcid.org/#1}
}

\newcommand{\I}{\bm{I}}
\newcommand{\argmin}{\mathop{\mathrm{arg~min}}\limits}

\begin{document}
\title{Gaussian Fourier Pyramid for Local Laplacian Filter}

\author{%
Yuto Sumiya,
Tomoki Otsuka,\\
Yoshihiro Maeda\orcidAuthor{0000-0001-6919-637X}~\IEEEmembership{Member,~IEEE,} and
Norishige Fukushima\orcidAuthor{0000-0001-8320-6407}~\IEEEmembership{Member,~IEEE}\\
\thanks{Manuscript received April 19, 2005; revised August 26, 2015. This work was supported by JSPS KAKENHI (18K19813, 19K24368, 21H03465, 21K17768).}
\thanks{Y. Sumiya and N. Fukushima were with Nagoya Institute of Technology, Japan (Web: \url{https://fukushima.web.nitech.ac.jp/en/}).}
\thanks{T. Otsuka was with Nagoya Institute of Technology, Japan.}
\thanks{Y. Maeda was with Tokyo University of Science, Japan.}
}


 %The paper headers
\markboth{IEEE Signal Processing Letters,~Vol.~xx, No.~xx, August~2015}%
{Shell \MakeLowercase{\textit{et al.}}: Bare Demo of IEEEtran.cls for IEEE Journals}



\maketitle

\begin{abstract}
Multi-scale processing is essential in image processing and computer graphics.
Halos are a central issue in multi-scale processing.
Several edge-preserving decompositions resolve halos, e.g., local Laplacian filtering (LLF), by extending the Laplacian pyramid to have an edge-preserving property.
Its processing is costly; thus, an approximated acceleration of fast LLF was proposed to linearly interpolate multiple Laplacian pyramids.
This paper further improves the accuracy by Fourier series expansion, named Fourier LLF.
Our results showed that Fourier LLF has a higher accuracy for the same number of pyramids.
Moreover, Fourier LLF exhibits parameter-adaptive property for content-adaptive filtering.
The code is available at:  \url{https://norishigefukushima.github.io/GaussianFourierPyramid/}.
\end{abstract}

\begin{IEEEkeywords}
Local Laplacian filter, Laplacian pyramid, Fourier series expansion, scale-space
\end{IEEEkeywords}

% \fi
\IEEEpeerreviewmaketitle

\section{Introduction}
\IEEEPARstart{M}{ulti-scale} processing is an essential tool for detail manipulation of images and is used for various detail and contrast enhancement: high-dynamic-range imaging~\cite{durand2002fast}, detail enhancement~\cite{farbman2008edge}, and contrast enhancement~\cite{dippel2002multiscale}.
Multi-scale processing decomposes an input image into multiple layers using Gaussian and Laplacian pyramids~\cite{adelson1984pyramid}, wavelet transform~\cite{mallat1999wavelet}, and difference of Gaussians for scale-space analysis~\cite{lindeberg2013scale}.

Unsharp masking is one of the simplest multi-scale methods; it consists of two layers: an input and a detail layer.
The detail layer is a subtraction of the blurred image from the input image.
This simple method produces large halos in steep edges; thus, nonlinear enhancement is used for coefficients to suppress large amplitudes.

The halo problem is resolved by edge-preserving multi-scale decomposition.
Instead of linear filtering, bilateral filtering~\cite{tomasi1998bilateral} is used to generate a base layer~\cite{durand2002fast}.
The bilateral filter is accelerated through fast Fourier transform (FFT)~\cite{durand2002fast}, and is further accelerated by state-of-the-art methods~\cite{chaudhury2013acceleration,sugimoto2019200fps,sumiya2020extending}.
Next, iterative bilateral filtering is extended to decompose multiple layers~\cite{fattal2007multiscale}; however, it is difficult to determine the parameters in bilateral filtering, which still suffers from halo artifacts.
Moreover, multi-scale decomposition is represented by other filters, such as least squares~\cite{farbman2008edge} and iterative local linear regression filtering~\cite{gu2012local}, which is similar to guide image filtering~\cite{he2013guided}.
Wavelet-based methods are also proposed for edge-preserving decomposition.
Edge-avoiding wavelets~\cite{fattal2009edge} construct a basis according to the edge content of the images.
Iterative bilateral filtering-based decomposition~\cite{fattal2007multiscale} is one of the \`{A}-Trous wavelet methods~\cite{holschneider1990real}.
The method is extended to multilateral filtering for the computer graphics context~\cite{dammertz2010edge}.
These approaches generate detailed signals and then manipulate the signals through remap functions: linear-tone curve, gamma curve, and S-tone curve.

Local Laplacian filtering (LLF)~\cite{paris2011local} manipulates the contrast of the input signals using remap functions and then generates detailed signals.
LLF locally enhances the image contrast and constructs a Laplacian pyramid for each pixel; thus, LLF can reduce halos with clearer images than the other multi-scale methods.
Parameter adaptation for the remap function further improves the quality.

The per-pixel construction of the Laplacian pyramid is costly; thus, an accelerated method is proposed, called fast LLF~\cite{aubry2014fast}.
Fast LLF reduces the number of Laplacian pyramids by selecting at finite sampled points.
Then, the pyramids are linearly interpolated.
However, when the number of sample points is insufficient, the approximation accuracy is low.
Moreover, the parameter-adaptive version of fast LLF~\cite{AdaptiveFastLLF2019} is not an approximation of na\"ive LLF since the precomputed pyramids are based on a fixed parameter.

For function approximations, Fourier series expansion is a better interpolation method than the linear one.
Therefore, this study proposes an approximation for LLF using Fourier series expansion, named \textit{Fourier LLF}.
Fourier LLF generates Fourier pyramids, which include cosine and sine pyramids, and then product-sums the pyramids.
Fourier LLF improves the accuracy and produces parameter-adaptive functionality.
The generated Fourier pyramids are independent of the remap function; thus, we can change the function by switching only the coefficient for the pyramids.
Once the pyramids are generated, we can change the parameter by $O(1)$ operation.

The main contributions of this study are as follows:
\begin{itemize}
  \item We formulate remap functions in LLF using Fourier series expansion, which has higher accuracy than that of the fast LLF~\cite{aubry2014fast}.
  \item We verify that Fourier LLF exhibits parameter-adaptive property for the remap function.
\end{itemize}

\section{Preliminary}
\subsection{Gaussian Pyramid and Laplacian Pyramid}
We introduce Gaussian and Laplacian pyramids, which are the bases of LLF.
The Laplacian pyramid is used for multi-scale processing and analysis of images for compression~\cite{burt1983laplacian}, texture synthesis~\cite{heeger1995pyramid}, and harmonization~\cite{sunkavalli2010multi}.
The traditional Laplacian pyramid processing directly enhances the coefficients of the pyramid~\cite{vuylsteke1994multiscale,fattal2007multiscale,li2005compressing}.

First, the Gaussian pyramid is introduced.
Let $\I:\mathcal{S}\mapsto\mathcal{R}$ be a $D$-dimensional $R$-tone input grayscale image, where
$\mathcal{S}\subset\mathbb{Z}^D$ and $\mathcal{R}\subset\mathbb{R}$ denote the spatial and range domain (generally, $D\!=\!2$ and $R\!=\!256$), respectively.
The Gaussian pyramid is defined as the set of $G_\ell[\I]\in\mathbb{R}$, where $\ell \in \mathcal{L} = \{0, 1,2,\ldots,\ell_{\max}\}\subset\mathbb{Z}$.
The lowest level of the pyramid is $G_0[\I]=\I$, and its other level $G_{\ell+1}[\I]$ is the downsampled blurred image of $G_\ell[\I]$. 
This relationship can be described as follows:
\vspace{-0.3em}
\begin{align}
G_{\ell}[\I] =  (\mathcal{G}_{\sigma}*G_{\ell-1}[\I])_{\downarrow},
\end{align}
where $\mathcal{G}_{\sigma}*$ is a Gaussian convolution with the standard deviation $\sigma$. 
$\downarrow$ is a downsampling operator, which halves the image size.
The size of $G_{\ell+1}[\I]$ is the half-width and height of $G_{\ell}[\I]$.
Usually, the Gaussian convolution is based on the binomial distribution for integer operations with five taps.
The regarded standard deviation is approximately $\sigma=1$.

The Laplacian pyramid is defined by the difference between the Gaussian pyramids of successive levels:
\vspace{-0.3em}
\begin{align}
L_\ell[\I] = G_{\ell}[\I] - (G_{\ell+1}[\I])_{\uparrow},
\end{align}
where $\uparrow$ is an upsampling operator that doubles the image width and height using a smoothing kernel.
Usually, the kernel is identical to $\mathcal{G}_{\sigma}$.
The highest level of the Laplacian pyramid is $L_{\ell_{\max}}[\I]=G_{\ell_{\max}}[\I]$.

\subsection{Manipulation of the Laplacian Pyramid}
The multi-scale detail enhancement with the Laplacian pyramid amplifies the pyramid coefficients except for the coarsest layer $\sup{\mathcal{L}}$.
For enhance functions, an S-tone-like function can suppress an overshoot, while the straightforward function is $r(i,0)=mi$, where $m\in\mathbb{R}$ is a constant value.
If $m=1$, the resulting image is the input image.
The argument $0$ in $r(i,0)$ is unnecessary in this case; it is used to match the latter remap function for LLF.
Finally, the remapped signals are collapsed to obtain the output:
\vspace{-0.2em}
\begin{align}
    \bm{O} = L_{\ell_{\max}}[\I] + \sum_{\ell \in \mathcal{L} \backslash \{\ell_{\max}\}}r(L_\ell[\I],0),
    \label{eq:output-original-LLF}
\end{align}
where $\backslash$ indicates the excluding operator from the set.

\subsection{Local Laplacian Filtering}
The output pixel value of LLF $\bm{O}_{p}$ is defined by the Laplacian pyramid $L[\bm{O}]_{p}$ at $p$ called the local Laplacian pyramid, where $p = (x,y)\in\mathcal{S}$ is the pixel position.
Collapsing the pyramid generates the output image:
\vspace{-0.2em}
\begin{align}
\label{eq:collapsllf}
  \bm{O}_{p}=\sum_{\ell\in \mathcal{L}}L_{\ell}[\bm{O}]_{p}.
\end{align}
Let us consider the procedure for calculating $L_{\ell}[\bm{O}]_{p}$.
\begin{enumerate}
  \item Repeat 2)-- 4) for all the pixels $p$ and levels $\ell$.
  \item Build the Gaussian pyramid $G_{\ell}[\I]$ at level $\ell$.
  \item Apply a remap function discussed later and create a contrast-transformed image $\tilde{\I}=r(I,G_{\ell}[\I]_{p})$.
  \item Construct a Laplacian pyramid for $\tilde{\I}$, $L_{\ell}[\tilde{\I}]$, and copy the pyramid coefficient at $p$, $L_{\ell}[\tilde{\I}]_{p}$, as an output Laplacian pyramid of $L_{\ell}[\bm{O}]_{p}$. The remapping value changes according to the level $\ell$ at $p$.
  \item Collapse the pyramid using Eq.~\eqref{eq:collapsllf} to obtain $\bm{O}$.
\end{enumerate}
The flow requires per-pixel and per-level construction of the pyramid; thus, its order is $O(N|\mathcal{L}|\cdot N\log|\mathcal{L}|)$, where $N$ and $|\mathcal{L}|$ are the number of pixels and levels, respectively. 
$O(N\log|\mathcal{L}|)$ is the order of the pyramid building.

\subsection{Fast Local Laplacian Filtering}
The na\"ive implementation accesses all layers of the pyramid for all pixels, while fast LLF~\cite{aubry2014fast} requires a limited number of pyramids for approximation.
The algorithm of fast LLF is defined as follows:
\begin{enumerate}
  \item Build the Gaussian pyramid $G_{\ell}[\I] \quad \forall \ell \in \mathcal{L}$. 
  \item Sample the intensities $k\!\in\!\mathcal{I} \!\subset\!\mathcal{R}$ at equal intervals $\tau$.
  \item Compute the remap images $\bm{R}_k\!=\!r(\I,\!\tau k)$ for each $k$ and build their Laplacian pyramids $L_{\ell}[\bm{R}_k]$.
  \item For each pyramid and pixels ($\ell \in \mathcal{L}$, $p \in \mathcal{S}$):
  \begin{enumerate}
    \item Obtain the pixel value $G_{\ell}[\I]_p$ in the Gaussian pyramid.
    \item Find $a (0 \leq a \leq 1) $ such that \\$ G_{\ell}[\I]_p = \tau((1-a)k+a(k+1)), j = \lfloor G_{\ell}[\I]_p \rfloor /\tau $.
    \item Compute the output pyramid by linearly interpolating the precomputed pyramids:\\${\qquad}L_{\ell}[\bm{O}]_p=(1-a)L_{\ell}[\bm{R}_{k}]_p+ aL_{\ell}[\bm{R}_{k+1}]_p$.
  \end{enumerate}
  \item Collapse the pyramid using \eqref{eq:collapsllf} to obtain $\bm{O}_p$.
\end{enumerate}
The flow requires $|\mathcal{I}|+1$ pyramids, where $|\mathcal{I}|$ is the number of sampled intensities. The order is $O ((|\mathcal{I}|+1) \cdot N\log|\mathcal{L}|)$.

\subsection{Remap Function}
\begin{figure}[t]
\centering
\subfigure[smooth ($m<0$)]{\includegraphics[width=0.35\columnwidth]{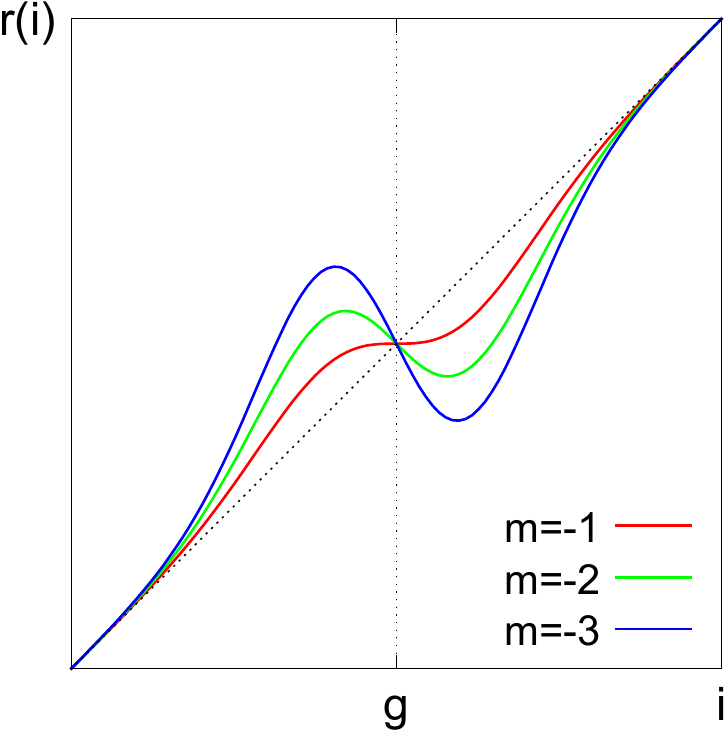} }
\subfigure[enhance ($m>0$)]{\includegraphics[width=0.35\columnwidth]{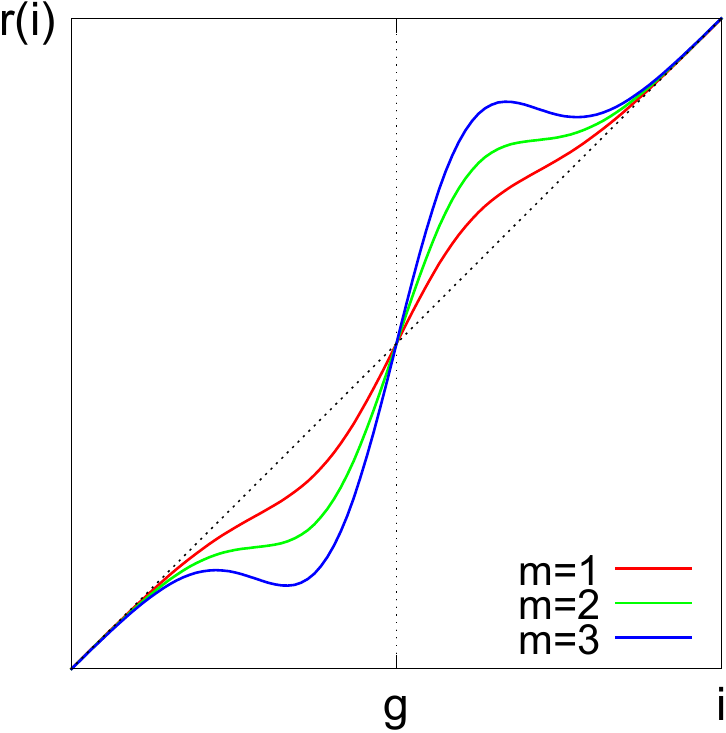} }
\vspace{-3mm}
\caption{\label{fig:remap}
Remap functions of detail smoothing and enhancement.
$\sigma_r$ is used to distinguish the details and edges.}
\vspace{-4mm}
\end{figure}

The remap function for an intensity $i \in \mathcal{R}$ is continuous and changes the input $i$ around a reference value $g \in \mathcal{R}$:
\vspace{-1mm}
\begin{align}
  \label{eq:remapping-function}
  r(i,g) = i - (i - g)f(i - g).
\end{align}
LLF uses $g= G_{\ell}[I]_p$.
$f\in\mathbb{R}$ is an even function usually and is specifically defined for each application.
This study uses a Gaussian function with multiplication factor $m$ denoted as $w_r(i)$ for $f$, which is defined as: 
%The $\sigma_r$ is scaled by 3.0 since the tail of the distribution is almost zero when $i=\sigma_r$:
\begin{align}
  \label{eq:remapping-function2}
  w_r(i) = m\exp\left(\frac{i^2}{-2\sigma_r^2}\right).
\end{align}
When $m>0$, the function enhances images, while when $m<0$, the function smooths images, as shown in Fig.~\ref{fig:remap}.
The dashed lines indicate the $y=x$ line, which is no remap function.
The difference between the $y=x$ line and the desired remap function indicates the degree of emphasis.

\section{Proposed Method}
\subsection{Formulation}
This study approximates the remap function using Fourier series expansion.
The derivative function of the Gaussian function $w_r$ in the remap function is:
\vspace{-1mm}
\begin{eqnarray}
  w^{'}_{r}(i-g) = \sigma^{-2}_{r}(i-g) w_{r}(i-g)\nonumber \\
  \therefore (i-g) w_{r}(i-g) = \sigma^{2}_{r} w^{'}_{r}(i-g).
  \label{diff_approx}
\end{eqnarray}
Substituting \eqref{diff_approx} into the remap function \eqref{eq:remapping-function}, the form is:
\vspace{-1mm}
\begin{eqnarray}
  r(i,g) &=& i - (i-g) w_{r}(i-g)\nonumber\\
  &=&i-\sigma^{2}_{r} w^{'}_{r}(i-g).\label{eq:RemapSubstitute}
\end{eqnarray}
The Gaussian function in $w_r(i)$ is an even function, which can be approximated by the number of finite cosine terms $K$ of the Fourier series expansion~\cite{sugimoto2015compressive}:
\vspace{-0.3em}
\begin{eqnarray}
  w_r(i-g) \approx m(\alpha_{0} + 2\sum_{k=1}^{K}\alpha_{k}\cos(\omega_{k}(i-g))),
  \label{approx_gaussF}
\end{eqnarray}
where, $\alpha_{k} = \frac{\sigma_r\sqrt{2\pi}}{T} \exp{(-\frac{1}{2}(\omega_k\sigma_{r})^{2})}$ and $\omega_{k} = \frac{2\pi}{T}k$.
$T$ is the period.
The derivative function of~\eqref{approx_gaussF} is 
\begin{eqnarray}
  w_{r}^{'}(i-g)\approx -2m\sum_{k=1}^{K}\alpha_{k}\sin(\omega_{k}(i-g))\omega_{k}.
  \label{approx_gauss}
\end{eqnarray}
Using~\eqref{approx_gauss} and the addition theorem of trigonometric functions, we can approximate the remap function \eqref{eq:RemapSubstitute} as follows:
{
\small
\begin{eqnarray}
r(i,g)\!\approx \!i\!-m\!\sum_{k=1}^{K}\!\tilde{\alpha_k}\left( \sin(\omega_{k}g)\cos(\omega_{k}i)\!-\!\cos(\omega_{k}g)\sin(\omega_{k}i) \right),\label{remap func}
\end{eqnarray}
}where $\tilde{\alpha}_k=2\sigma_{r}^{2}\alpha_{k}\omega_{k}$.
%The technique was also used for approximating Gaussian kernel in the bilateral filter~\cite{deng2017fast}.

The period $T$ for an $R$-tone image is usually determined such that the following equation becomes minimum~\cite{sugimoto2015compressive}; the equation approximates the formula for the error between the Gaussian function and its approximation:
\begin{align}
\!\!\argmin_{T} E_k(T)\!=\! \mathrm{erfc} \left(\frac{\pi \sigma}{T} (2K\!+\!1)  \right) \!+\! \mathrm{erfc} \left(\frac{T\!-\!R}{\sigma} \right).
\end{align}

\begin{figure}[t]
\centering
\includegraphics[width = 0.9\columnwidth]{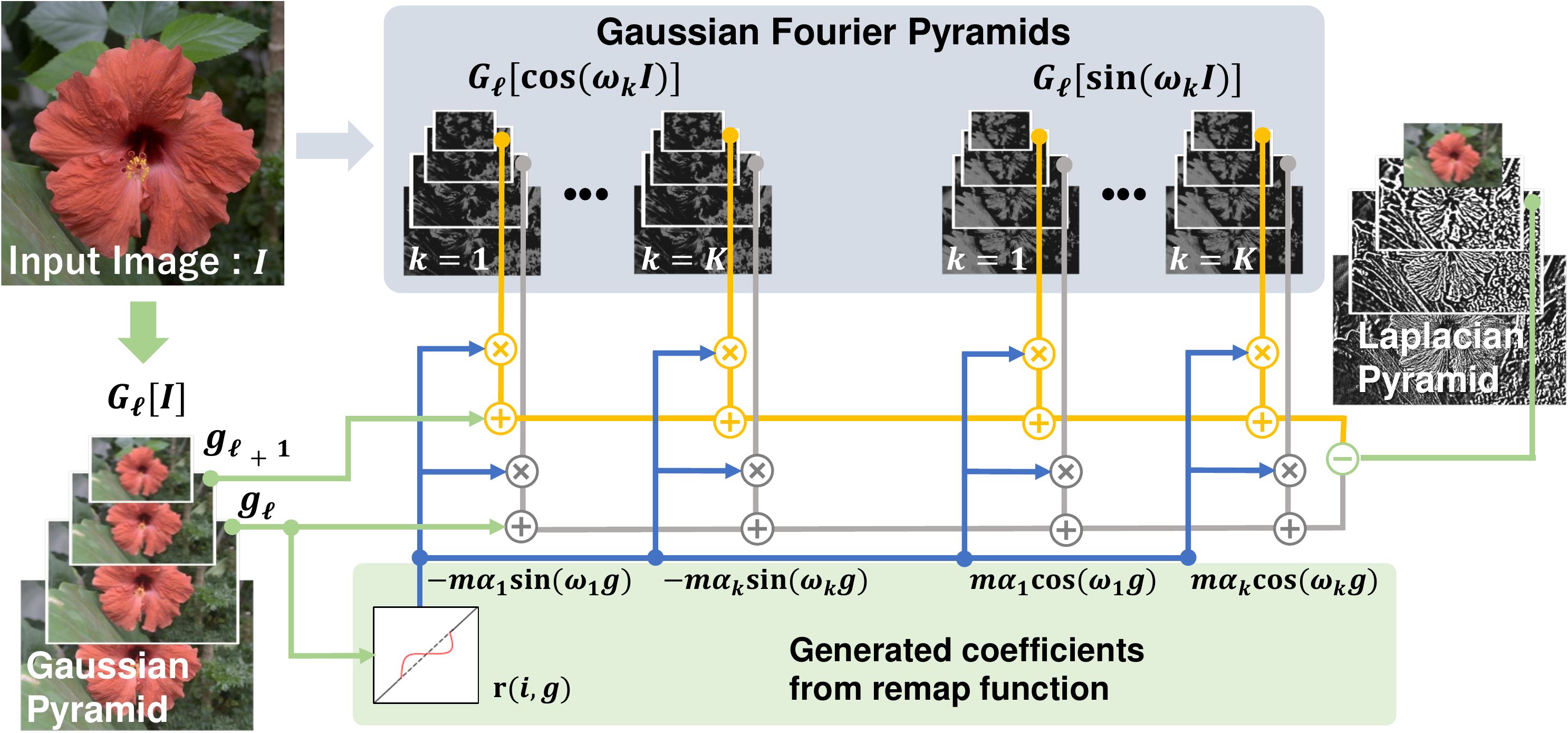}
\vspace{-3mm}
\caption{
Overview of the Fourier LLF, which shows the computational flow of ~\eqref{eq:proposed_eq}.
First, we create a Gaussian pyramid of the input image ($G_{\ell}[I]$) and Gaussian Fourier pyramids ($G_{\ell}[\sin(\omega_{k}I)], G_{\ell}[\cos(\omega_{k}I)]$).
Next, we compute $m\alpha_k\cos(\omega_{k}g)$ and $m\alpha_k\sin(\omega_{k}g)$ according to the value of the Gaussian pyramid $g$, and the remap function.
The output Laplacian pyramid is approximated by the product-sum of the Gaussian Fourier pyramids and coefficients, except the top.
%This process is done $K$ times for every pixel in every level except the top one.
}
\label{fig:1}
\vspace{-3mm}
\end{figure}

The output local Laplacian pyramid is defined by
\begin{eqnarray}
  L_{\ell}[\bm{O}]_{p}\!=\!G_{\ell}[r(\I,G_{\ell}[\I]_{p})]\!-\!G_{\ell+1}[r(\I,G_{\ell}[\I]_{p})]_{\uparrow}.
  \label{LLF_def}
\end{eqnarray}
Recall that $G_0[\I]=\I$, $L_{\ell_{\max}}[\bm{O}]=G_{\ell_{\max}}[\I]$, and $r(\I,\I_p)=\I$; thus, the form is expressed as follows by substituting~\eqref{remap func}:
{\footnotesize
\begin{align}
  &\!\rm{case}: \ell=0\nonumber\\
  &L_{0}[\bm{O}] \!\approx\! \I\!-\!G_{1}[\I]_{\uparrow}\!+\!m\!\sum_{k=1}^{K}\!\tilde{\alpha_k}(\sin(\omega_{k}G_{0}[\I])\tilde{C}_{{1,k}_{\uparrow}}\!-\!\cos(\omega_{k}G_{0}[\I])\tilde{S}_{{1,k}_{\uparrow}}),\nonumber\\
  &\!\rm{case}: 1 \leq \ell \leq \ell_{\max}-1\nonumber\\
  &\,\,\,L_{\ell}[\bm{O}] \approx G_{\ell}[\I]-m\sum^{K}_{k=1}\tilde{\alpha_k}(\sin(\omega_{k}G_{\ell}[\I])\tilde{C}_{\ell,k}\!-\!\cos(\omega_{k}G_{\ell}[\I])\tilde{S}_{\ell,k})\nonumber\\
  &\hspace{-1mm}-G_{\ell\!+\!1}[\I]_{\uparrow}+m\sum_{k=1}^{K}\tilde{\alpha_k}(\sin(\omega_{k}G_{\ell}[\I])\tilde{C}_{{\ell\!+\!1,k}_{\uparrow}}\!\!-\!\cos(\omega_{k}G_{\ell}[\I])\tilde{S}_{{\ell\!+\!1,k}_{\uparrow}}),\nonumber\\
  &\mbox{where }\tilde{C}_{\ell,k}=G_{\ell}[\cos(\omega_{k}\I)] \hspace{0.5em}\mbox{and}\hspace{0.5em} \tilde{S}_{\ell,k}=G_{\ell}[\sin(\omega_{k}\I)].
  \label{eq:proposed_eq}
\end{align}
}
Figure~\ref{fig:1} represents~\eqref{eq:proposed_eq}, visually.
$\tilde{C}_{\ell,k}$ and $\tilde{S}_{\ell,k}$ represent the $\ell$-th layer of the $k$-th Gaussian Fourier pyramids (blue shaded area), and $\cos{(\omega_{k}G_{\ell}[I])}$ and $\sin{(\omega_{k}G_{\ell}[I])}$ multiplied by $m\tilde{\alpha_k}$ are the coefficients (green shaded area).
In~\eqref{eq:proposed_eq}, all image operators (such as $L_{\ell}[\cdot], G_{\ell}[\cdot], \tilde{C}_{\ell,k}, \tilde{S}_{\ell,k})$ can be followed by the pixel position $p$.
We can remove the operator $p$ from~\eqref{LLF_def} because there is no loop dependency for $p$.
Because the arguments of all the pyramids are given by the input image and constant variables, the pyramids are independently computed; thus, the output image $O_p=\sum_{\ell \in \mathcal{L}}L_{\ell}[O]_{p}$ is obtained by the computing cost of $2K+1$ pyramids: $K\!$ cosine and $K\!$ sine pyramids, and $G_{\ell}[I]$.
Therefore, the computational order of our method is $O((2K+1)\cdot N\log|\mathcal{L}|)$.

\subsection{Pixel-by-Pixel Enhancement}
Changing the remap function for each pixel is essential for enhancement, such as the enhancement of flat regions to avoid  noise signal boosting.
The na\"ive implementation can locally change the function without any additional footprint; however, the computation itself is inefficient.
The paper~\cite{AdaptiveFastLLF2019} also uses adaptive remap functions for fast LLF; however, the implementation is not an approximation of LLF.
Fast LLF requires Laplacian pyramids for each remap function to interpolate the pyramid; thus, the multiple of the number of remap functions and pyramids is required for approximating LLF.

By contrast, the proposed method is independent of the remap functions for generating pyramids because the Gaussian Fourier pyramids of $\bm{I}$ do not depend on the remap function.
Only the coefficients exhibit a dependency on the remap function; thus, we change the coefficient for the pixel-level parameter adaptation.
Let $\tilde{\alpha}_{k,p}$ be the $k$-th coefficient for $p$. 
The remap function can be switched by replacing $\tilde{\alpha}_{k,p}$ pixel-by-pixel in~\eqref{eq:proposed_eq} instead of $\tilde{\alpha}_{k}$.
Similarly, we can also change the magnification factor of $m$ to $m_p$ for pixel-by-pixel adaptation.
Therefore, we need $2K+1$ pyramids for this case, which are identical to the parameter-fixed case.

\section{Experimental Results}
We compared Fourier LLF with na\"ive LLF~\cite{paris2011local} and fast LLF~\cite{aubry2014fast}.
The processing time was evaluated using two types of CPU: AMD Ryzen Threadripper 3970X (32 cores 3.7GHz) and Intel Core i9-9980XE (18 cores 3.0GHz).
The code is written in C++, parallelized by OpenMP, and vectorized by AVX.
We removed the subnormal numbers for an efficient implementation~\cite{maeda2018edge}.
For the entire experiment, we set $\sigma_{r}\!=\!30$ and $T\!=\!405.9$, which is the optimal value when $\sigma_{r}\!=\!30$.
We used 10 traditional grayscale images of size $512\!\times\!512$ as test images.
%, such as flower, stone building, motocross bikes, sailboat at anchor, shuttered windows, sailboats under spinnakers, mountain stream, lighthouse in Maine, P51 Mustang, Portland Head Light, and mountain chalet.

\begin{figure}[t]
\begin{center}
  \subfigure{
  \includegraphics[width=0.22\columnwidth]{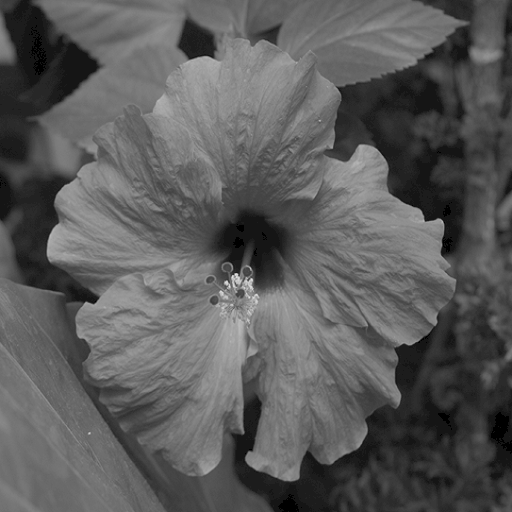}
  }
   \subfigure{
  \includegraphics[width=0.22\columnwidth]{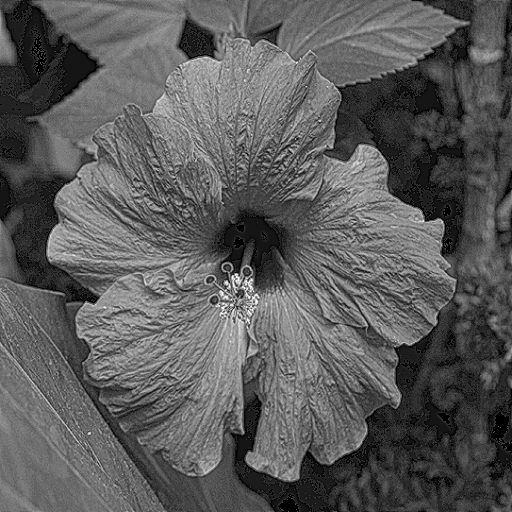}
  }
  \subfigure{
  \includegraphics[width=0.22\columnwidth]{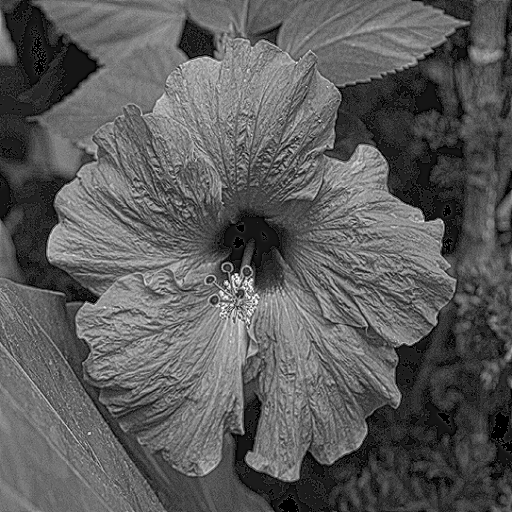}
  }\\
  \addtocounter{subfigure}{-3}
  \centering
  \vspace{-3mm}
  \subfigure[input]{
  \includegraphics[width=0.22\columnwidth]{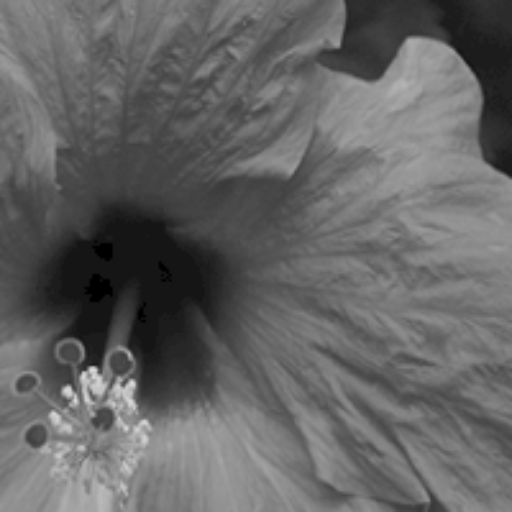}\label{inputImg}
  }
  \subfigure[fast LLF]{
  \includegraphics[width=0.22\columnwidth]{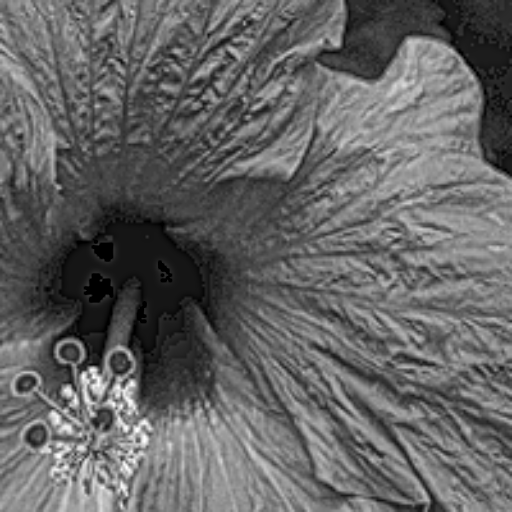}\label{fastLLFImg}
  }
  \subfigure[Fourier LLF]{
  \includegraphics[width=0.22\columnwidth]{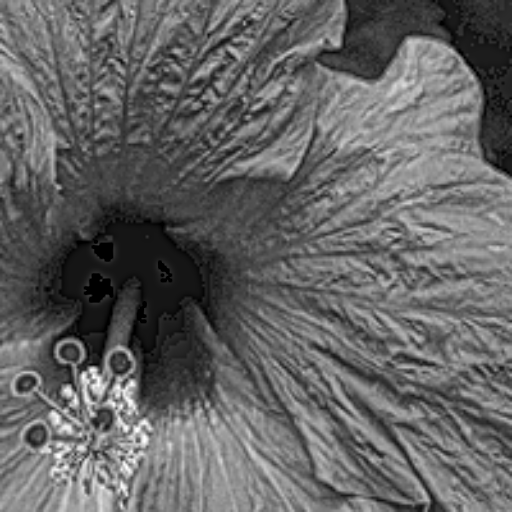}\label{proposedImg}
  }
  \end{center}
    \vspace{-4mm}
  \caption{Input and output images (2-layer and 21 pyramids).}
  \vspace{-2mm}
  \label{outputImgs}
\end{figure}
\begin{figure}[t]
\centering
\subfigure[PSNR]{
\includegraphics[width = 0.5\columnwidth]{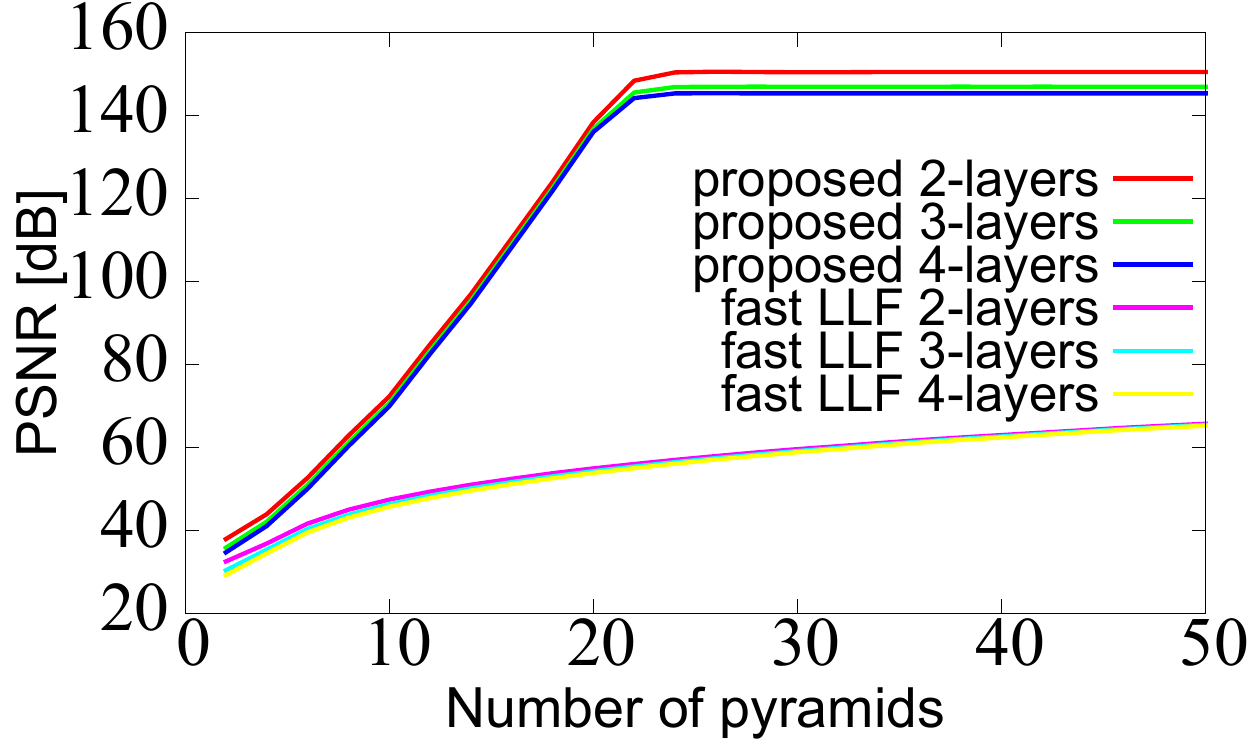}\label{all}
}
\subfigure[PSNR at 25 pyramids]{
\includegraphics[width = 0.4\columnwidth]{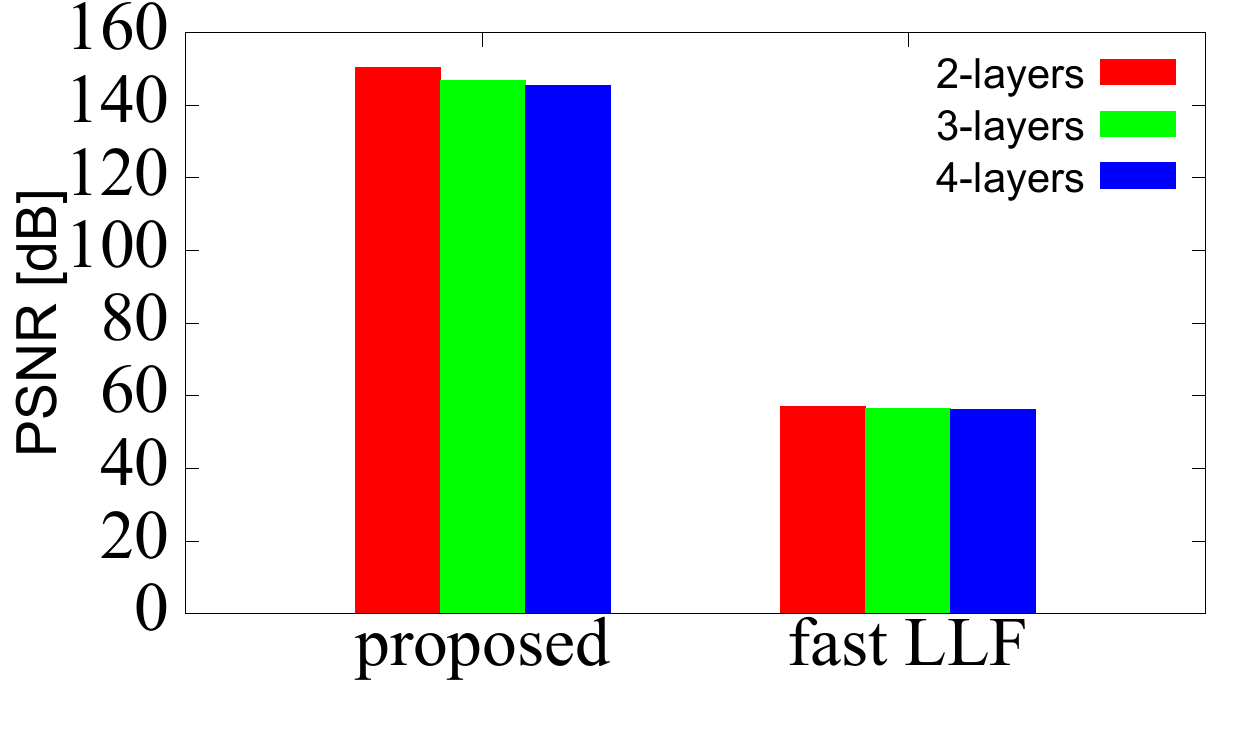}\label{box}
}
  \vspace{-2mm}
  \caption{(a) PSNR of Fourier LLF and fast LLF to the number of pyramids (average of 10 images). (b) PSNR when the number of pyramids is 25.  The ground truth is the na\"ive LLF output.}
  \label{psnrall}
\end{figure}
\begin{comment}
\begin{figure}[t]
    \subfigure[PSNR of 2 layers]{
  \includegraphics[width=0.46\columnwidth]{fig/PSNR_L1.pdf}\label{PSNR_2layers}
  }
  \subfigure[PSNR of 3 layers]{
  \includegraphics[width =0.46\columnwidth]{fig/PSNR_L2.pdf}\label{PSNR_3layers}
  }
  \vspace{-2mm}
  \caption{PSNR (average of 10 images) of Fourier LLF (proposed method) and fast LLF. The ground truth is the na\"ive LLF output.}
  \label{psnr}
\end{figure}
\end{comment}
Figure~\ref{outputImgs} shows the input and output images for fast LLF and Fourier LLF with 21 pyramids.
In both outputs, the details of the input image are emphasized.
The processing time of our method is 16.5ms, and that of fast LLF is 15.2ms in Threadripper. 
In Core i9, it is 20.2ms for our method and 21.3ms for fast LLF.
The processing time of na\"ive LLF is about 5 min in both CPUs.

These values are the average times of 100 trials.
The processing time difference between the CPUs is not significant.
When the number of pyramids is the same, the computation speed is almost the same.

%Figures~\ref{PSNR_2layers} and \ref{PSNR_3layers} show PSNR of fast LLF and our method for each 2-layers and 3-layers case.
%The x-axis shows the number of pyramids, which can be considered as the processing time since the building pyramid cost is dominant in the filtering.
%There is not much difference between the two methods up to about six pyramids, while beyond that, our method shows a higher PSNR value per pyramid.
%The processing time with the same pyramid numbers is almost the same; the proposed method is superior to fast LLF.

Figure~\ref{psnrall} shows PSNR of fast LLF and our method for the 2-layer, 3-layer, and 4-layer cases.
In Fig.~\ref{all}, the $x$-axis shows the number of pyramids, which can be considered as the processing time because the cost of building the pyramids is a dominant factor in the filtering.
For the same number of pyramids, our method always shows a higher PSNR than that of fast LLF.
The processing time with the same number of pyramids is almost the same; however, the overall performance of the proposed method is superior to that of fast LLF.
Figure~\ref{box} shows the PSNR values when the number of pyramids is 25.
It can be seen that for both methods, the PSNR decreases slightly as the number of layers increases, owing to accumulation errors in floating-point numbers.
Note that over 59 dB images have the same output in the 8-bit level and are consistent with the output of the original paper~\cite{paris2011local}.
Therefore, please see~\cite{paris2011local} for a visual comparison with other enhancement methods or our supplemental document.

\begin{figure}[t]
 \begin{center}
  \subfigure{
  \includegraphics[width=0.21\columnwidth]{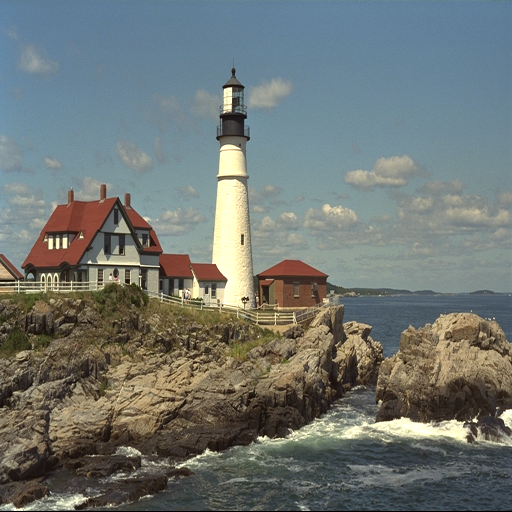}
  }
  \subfigure{
  \includegraphics[width=0.21\columnwidth]{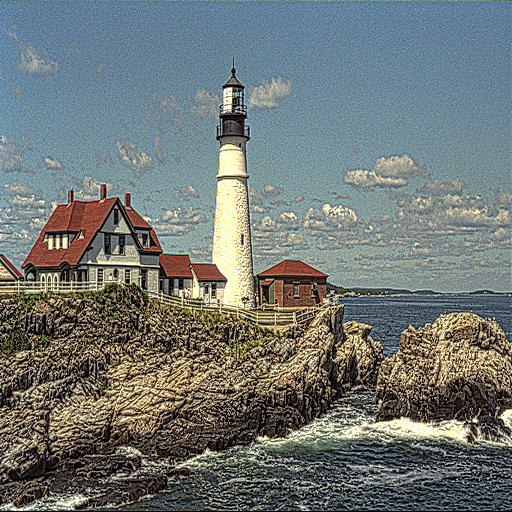}
  }
  \subfigure{
  \includegraphics[width=0.21\columnwidth]{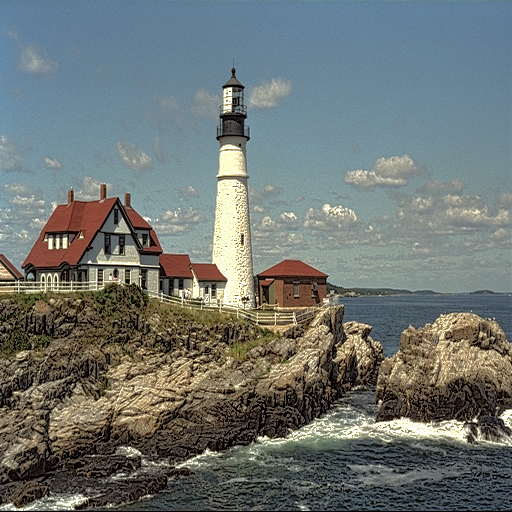}
  }\\
    \addtocounter{subfigure}{-3}
      \vspace{-3mm}
  \subfigure[input]{
  \includegraphics[width=0.21\columnwidth]{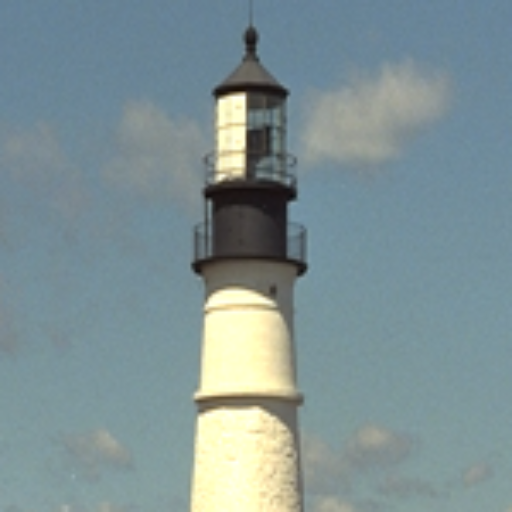}\label{InputColor}
  }
  \subfigure[fixed]{
  \includegraphics[width=0.21\columnwidth]{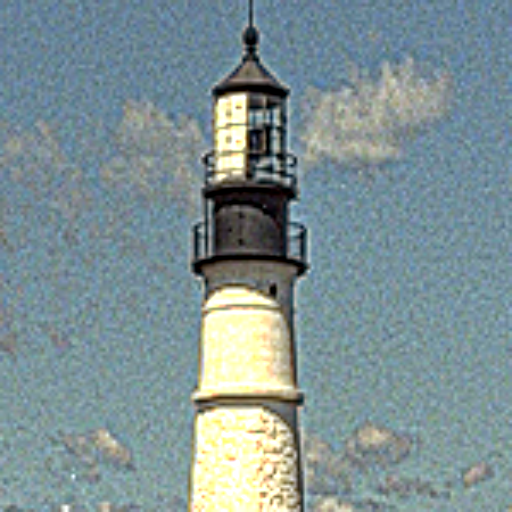}\label{SimpleLLF}
  }
  \subfigure[pix-by-pix]{
  \includegraphics[width=0.21\columnwidth]{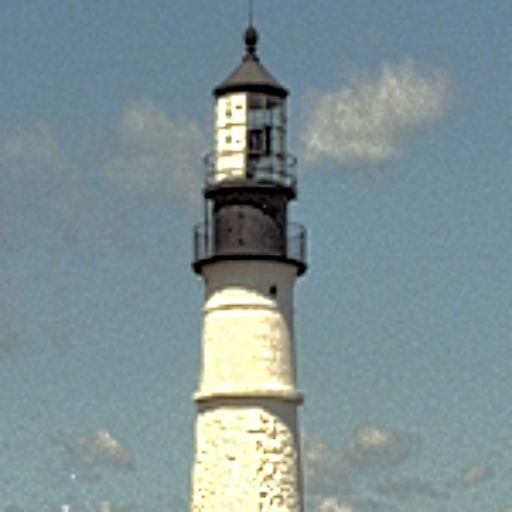}\label{PixelLLF}
  }
  \end{center}
  \vspace{-4mm}
  \caption{Outputs of pixel-by-pixel enhancement of Fourier LLF.}
      \label{fig:pixel_enhance_image}
    \vspace{-2mm}
\end{figure}

Figure~\ref{fig:pixel_enhance_image} demonstrates the pixel-by-pixel enhancement.
The degree of enhancement is determined according to the variance of the pixels in a local window; the low variance is a small enhancement.
In the fixed-parameter case (Fig.~\ref{SimpleLLF}), the lighthouse is emphasized, while the flat sky is also emphasized, which may degrade the quality of the image.
The pixel-by-pixel method (Fig.~\ref{PixelLLF}) can enhance the image while maintaining visual of flat areas.
For color processing, LLF is applied to the Y channel of the YUV color space, and the other components are kept.

\begin{figure}[t]
  \centering
  \includegraphics[width = 0.45\columnwidth]{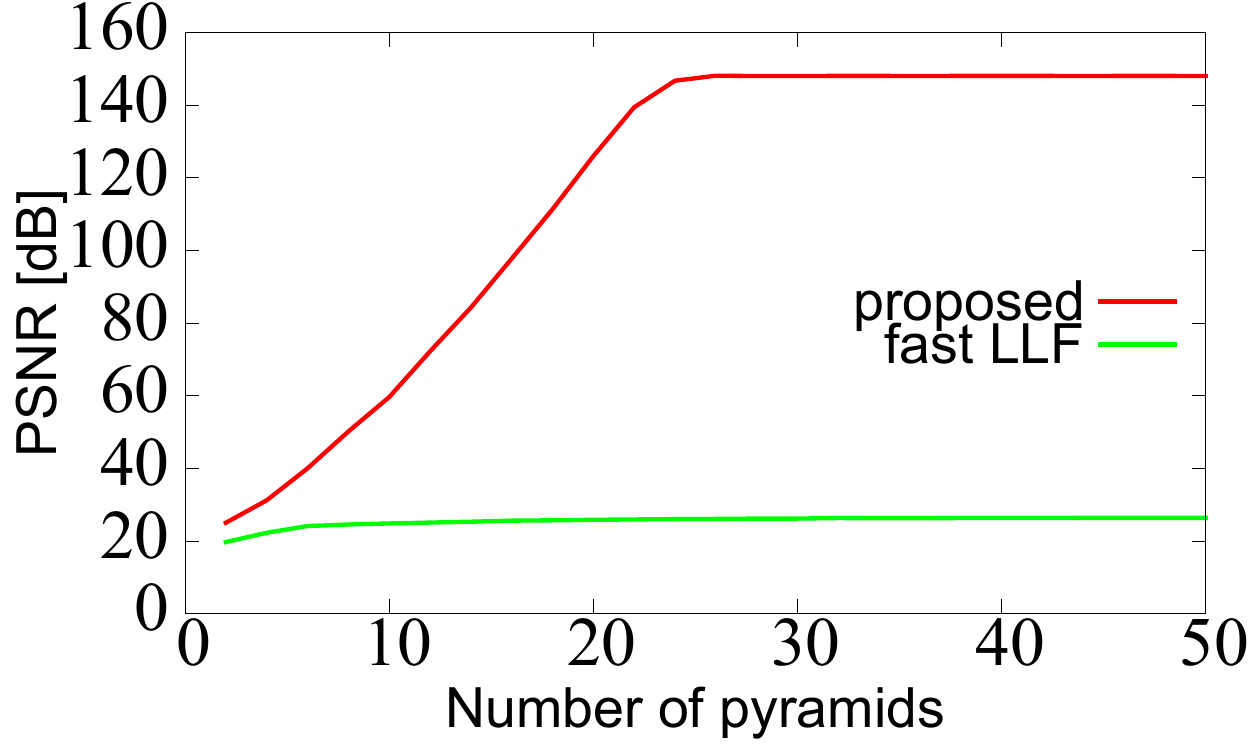}
  \vspace{-4mm}
  \caption{Pixel-by-pixel enhancement case: PSNR of the proposed method and adaptive fast LLF~\cite{AdaptiveFastLLF2019}. The ground truth is the pixel-by-pixel enhanced na\"ive LLF.}
  \label{fig:pixel_enhance_psnr}
\end{figure}

Figure~\ref{fig:pixel_enhance_psnr} shows PSNR of parameter-adaptive fast LLF~\cite{AdaptiveFastLLF2019} and Fourier LLF with the pixel-by-pixel enhancement.
Adaptive fast LLF does not improve PSNR even when the number of pyramids increases because fast LLF assumes that the same remap functions are used for all pixels.
Fast LLF with the adaptive remap function~\cite{AdaptiveFastLLF2019} is another filter from the parameter adaptive na\"ive LLF.
By contrast, the proposed method maintains high accuracy in the parameter-adaptive case.


\section{Conclusion}
This study proposes an approximation for LLF using Fourier series expansion, called Fourier LLF.
The experimental results showed that our method achieved higher accuracy per pyramid building and better performance than those of the conventional method.
We also showed that our method can approximate the adaptive filter.
The limitation of the proposed method is that it cannot handle color image distances.
Note that the conventional method is also for grayscale images.
The aforementioned limitation can be solved by high-dimensional Gaussian kernel approximation methods~\cite{nair2019fast,miyamura2020image}.

\section*{Supplemental material}
\begin{figure}[tb]
%\vspace{-10cm}
\centering
\subfigure[Input]{
\
  \includegraphics[width=0.45\columnwidth]{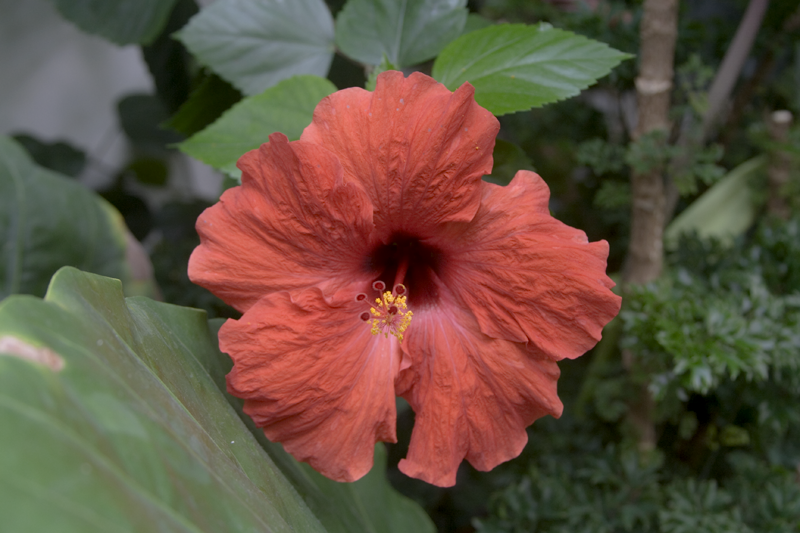}
  }
  \subfigure[2 layers]{
  \includegraphics[width=0.45\columnwidth]{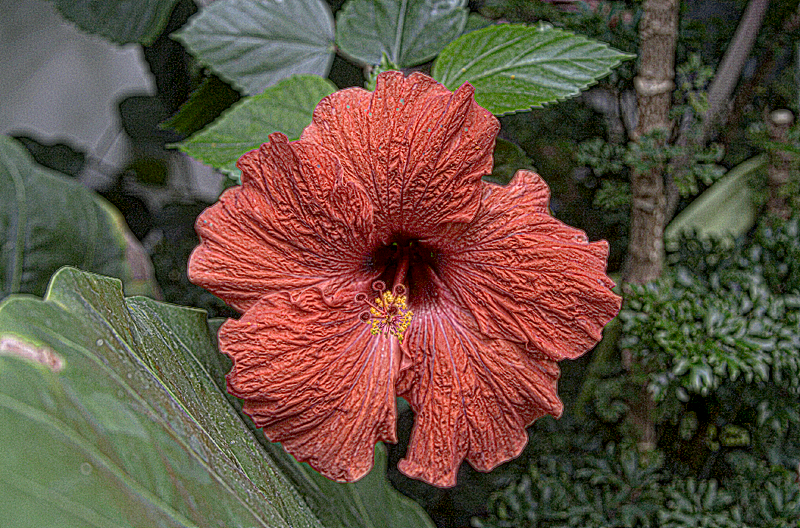}
  }\\\vspace{-2mm}
   \subfigure[3 layers]{
  \includegraphics[width=0.45 \columnwidth]{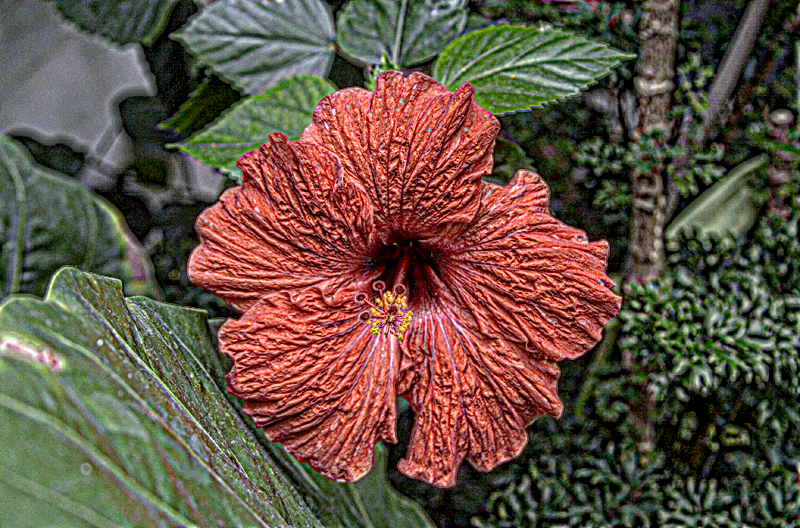}
  }
  \subfigure[4 layers]{
  \includegraphics[width=0.45 \columnwidth]{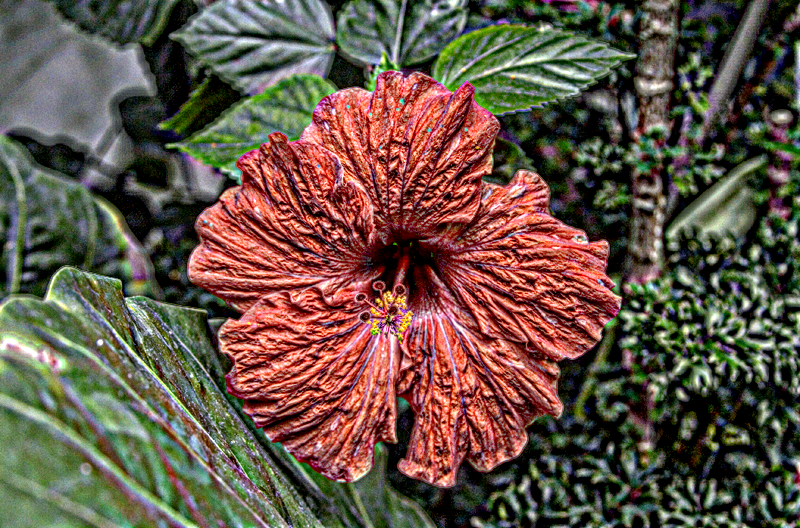}
  }
  \vspace{-2mm}
  \caption{Proposed LLF with various layers for Flower image ($\sigma_r = 15$, $m=7$, $K=10$).}
  \vspace{-5mm}
  \label{layer_flower}
\end{figure}

\begin{figure}[tb]
\centering
\subfigure[Input]{
  \includegraphics[width=0.45\columnwidth]{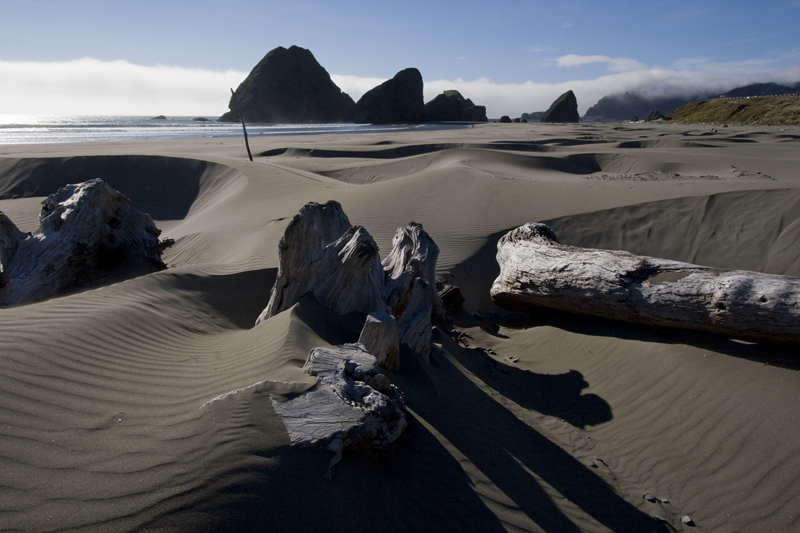}
  }
  \subfigure[2 layers]{
  \includegraphics[width=0.45\columnwidth]{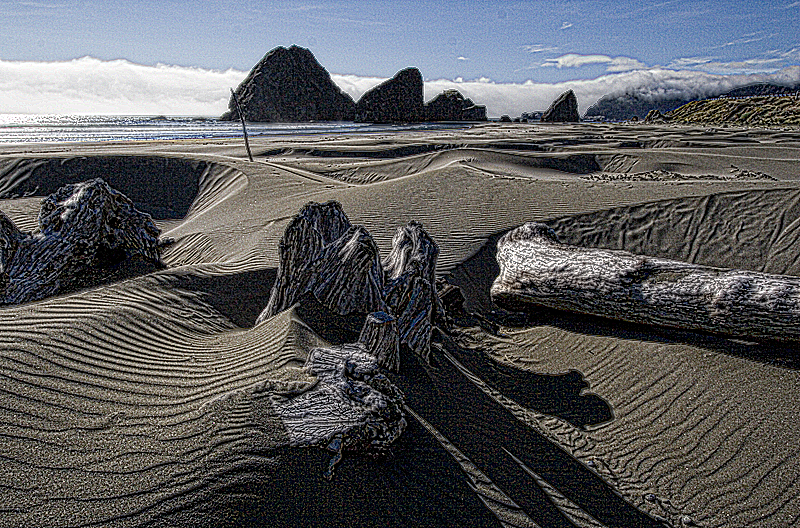}
  }\\\vspace{-2mm}
   \subfigure[3 layers]{
  \includegraphics[width=0.45 \columnwidth]{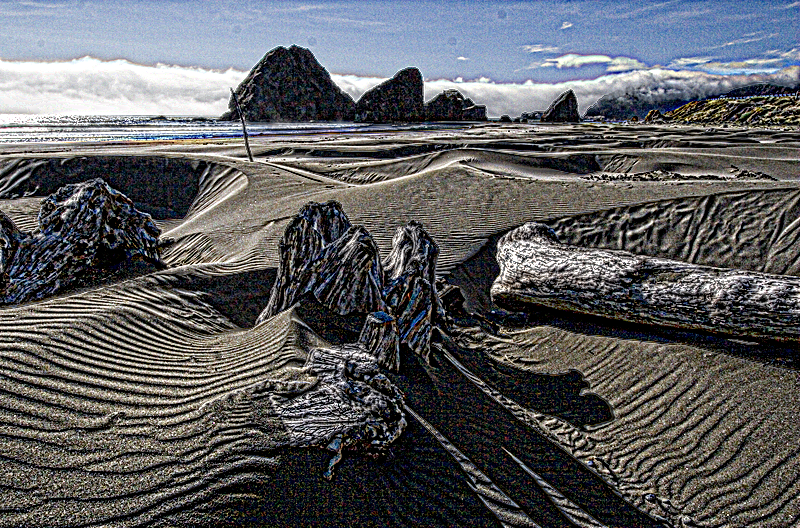}
  }
  \subfigure[4 layers]{
  \includegraphics[width=0.45 \columnwidth]{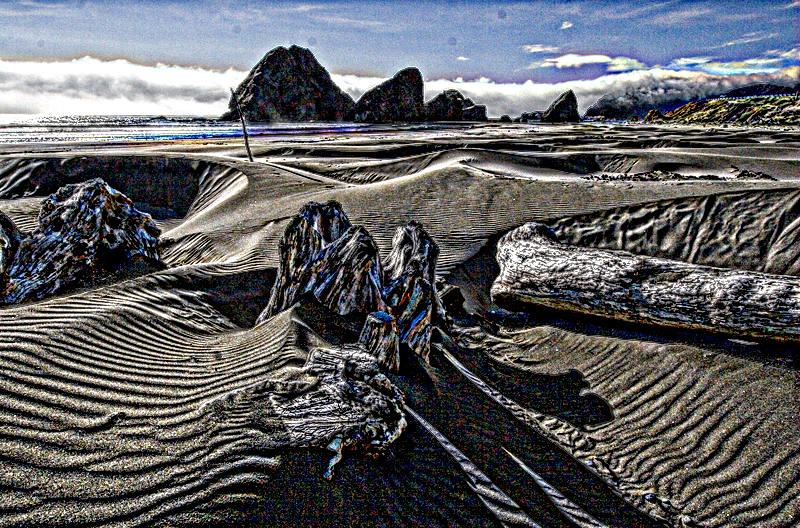}
  }
  \vspace{-2mm}
  \caption{Proposed LLF with various layers for Beach image ($\sigma_r = 15$, $m=7$, $K=10$).}
  \vspace{-5mm}
  \label{layer_beach}
\end{figure}

\begin{figure}[tb]
\centering
%\subfigure[Input]{
%  \includegraphics[width=0.45\columnwidth]{fig/flower.png}
%  }\\
  \subfigure[Proposed LLF]{
  \includegraphics[width=0.45\columnwidth]{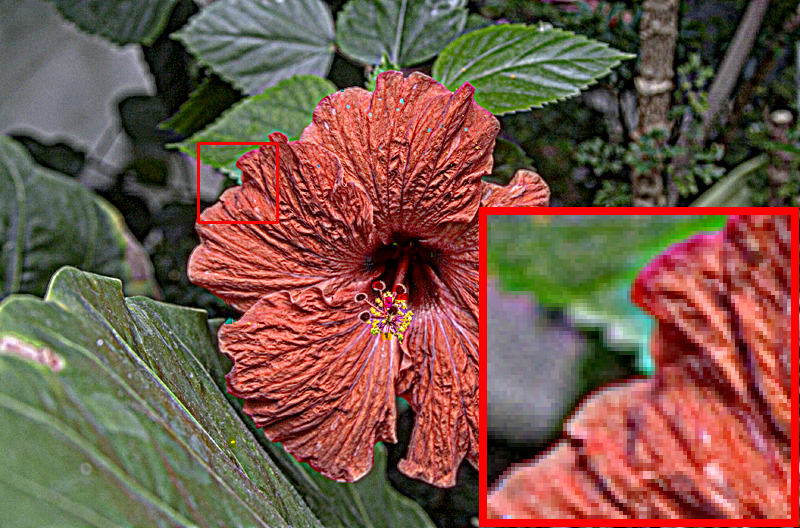}
  }
   \subfigure[Gauss]{
  \includegraphics[width=0.45\columnwidth]{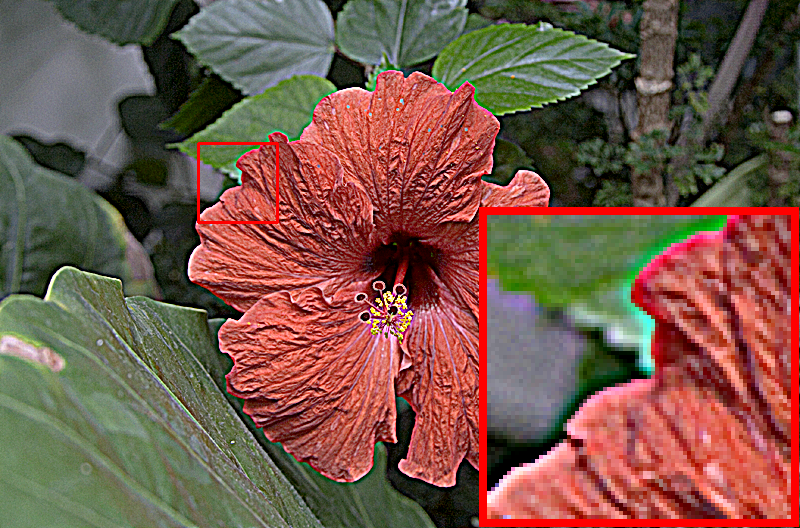}
  }
  \subfigure[Bilateral]{
  \includegraphics[width=0.45\columnwidth]{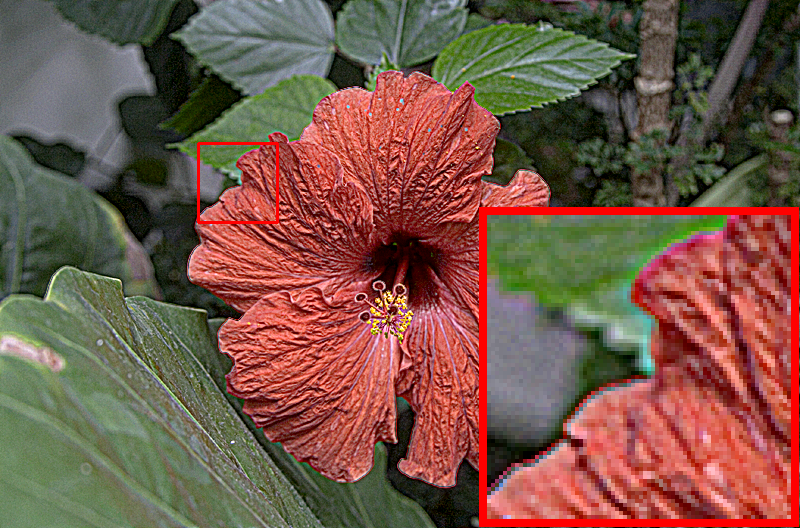}
  }
  \subfigure[Domain]{
  \includegraphics[width=0.45\columnwidth]{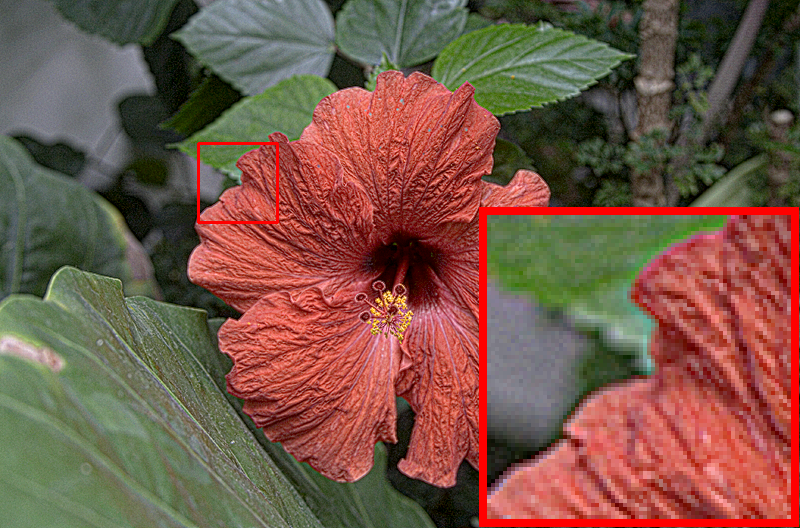}
  }
  \subfigure[Guided]{
  \includegraphics[width=0.45\columnwidth]{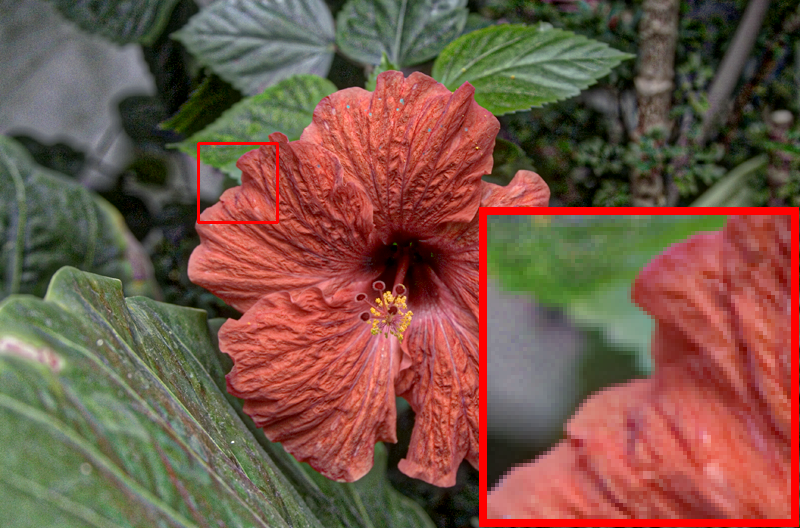}
  }
  \subfigure[Pyramid]{
  \includegraphics[width=0.45\columnwidth]{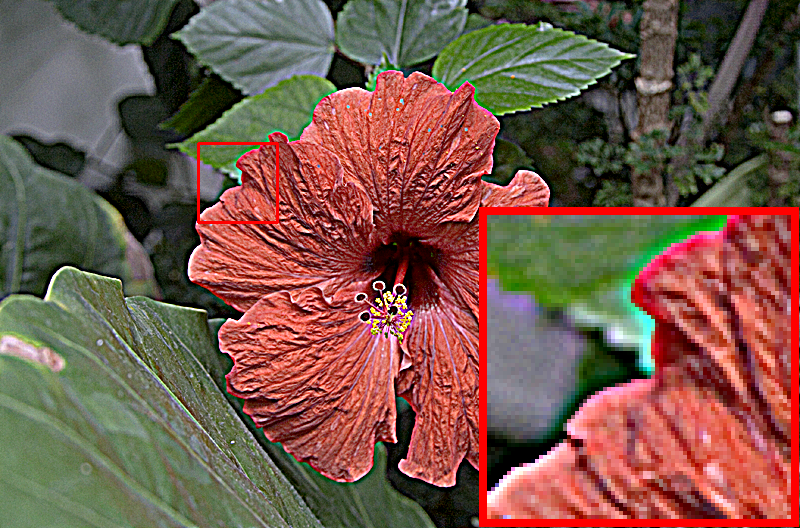}
  }
  \subfigure[MSBF]{
  \includegraphics[width=0.45\columnwidth]{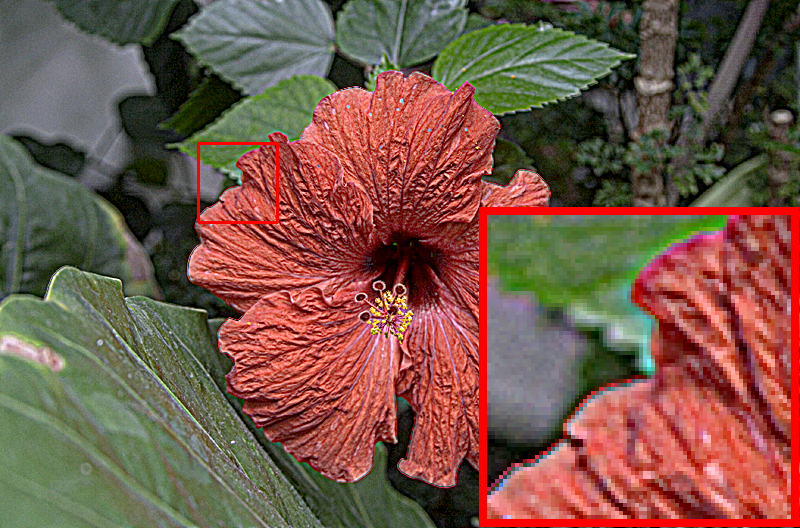}
  }
  \subfigure[EAW]{
  \includegraphics[width=0.45\columnwidth]{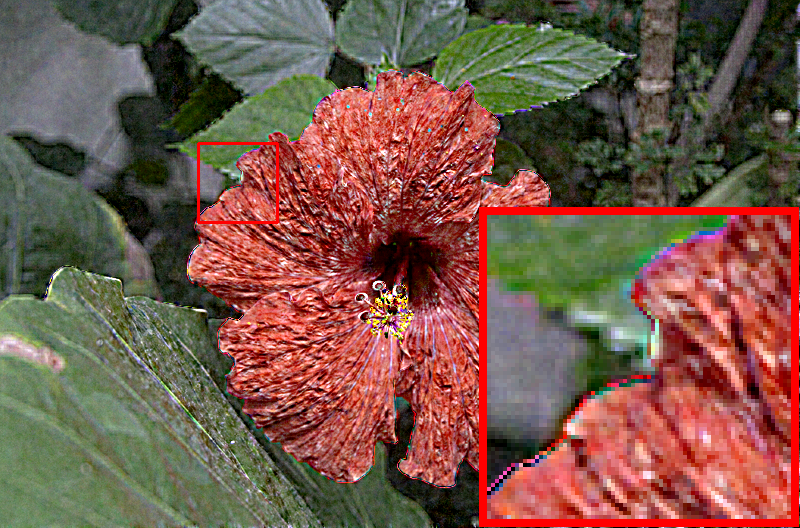}
  }
  \caption{Visual comparison with various method for Flower image (3-layers, $\sigma_r=30$, $m=7$, $K=10$).}
  \label{flower}
\end{figure}

We compared the Gaussian Fourier pyramid based LLF with two-layers enhancement methods and multi-layers enhancement methods.
The two-layers methods are defined as follows:
\begin{equation}
    \bm{O} = \I + r(\I-f*\I,0),
\end{equation}
where, $\I$ and $\bm{O}$ are input and output images, respectively.
$r$ is the remap function of Eq. (8).
$f*$ is a filter function, and this paper used four filters: Gaussian filter, bilateral filter~\cite{tomasi1998bilateral}, domain transform filter~\cite{GastalOliveira2011DomainTransform}, and guided image filter~\cite{he2013guided}.
The two-layer with Gaussian filtering means unsharp masking.
For multi-layers methods, we used three approaches: Laplacian pyramid enhancement of Eq. (3), multiscale bilateral filter (MSBF)~\cite{fattal2007multiscale}, and edge-avoiding wavelet (EAW)~\cite{fattal2009edge}.
For the experiments, we used enough order $K$ ($>$ 59dB). 
PSNR over 59dB means that the intensity difference between the ideal and approximated output is $<0.5$, generating the same integer value images.
Thus, the output of our LLF is the same as the original LLF~\cite{paris2011local}.

Figures~\ref{layer_flower} and \ref{layer_beach} show the effect of the multiple layers.
The LLF with more layers enhances images strongly. 

\begin{figure}[tb]
\centering
%\subfigure[Input]{
  %\includegraphics[width=0.45\columnwidth]{fig/beach.png}
  %}
  \subfigure[Proposed LLF]{
  \includegraphics[width=0.45\columnwidth]{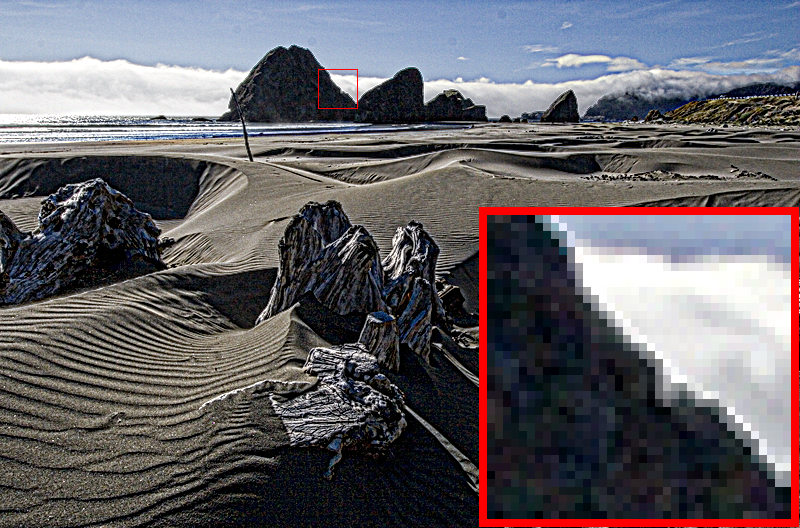}
  }
   \subfigure[Gauss]{
  \includegraphics[width=0.45\columnwidth]{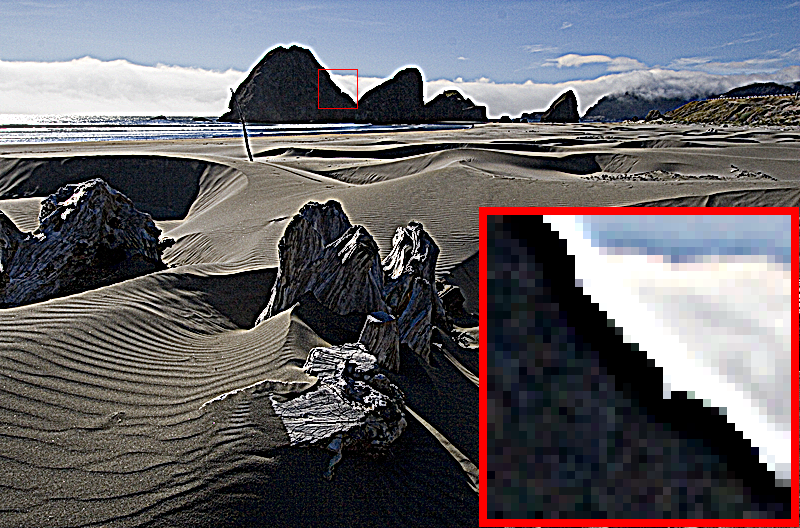}
  }
  \subfigure[Bilateral]{
  \includegraphics[width=0.45\columnwidth]{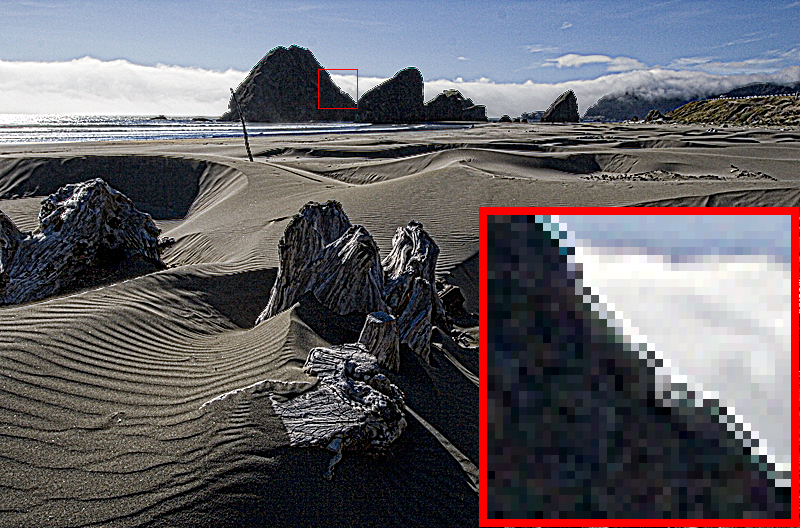}
  }
  \subfigure[Domain]{
  \includegraphics[width=0.45\columnwidth]{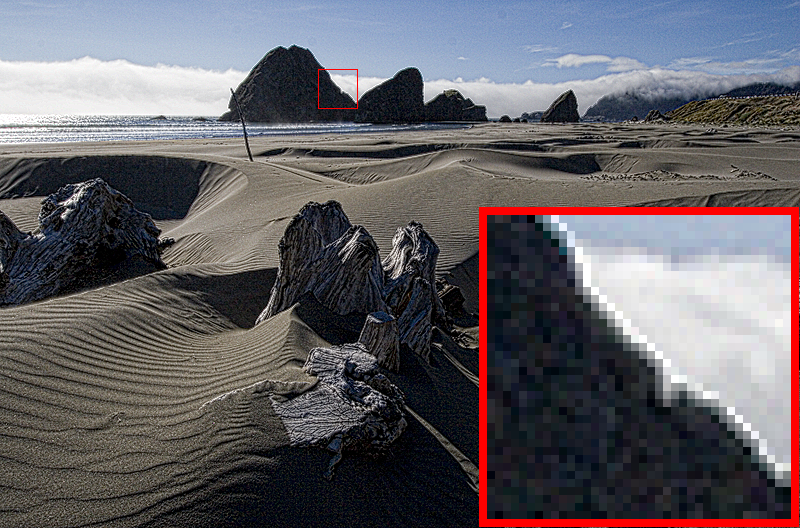}
  }
  \subfigure[Guided]{
  \includegraphics[width=0.45\columnwidth]{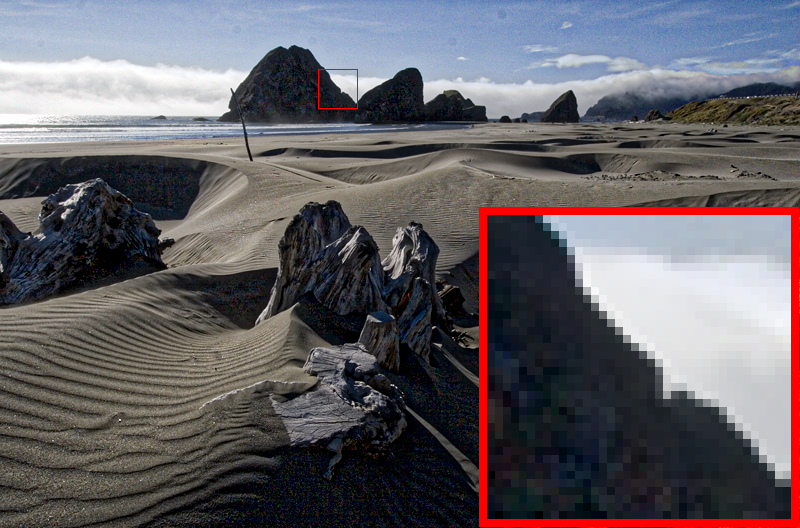}
  }
  \subfigure[Pyramid]{
  \includegraphics[width=0.45\columnwidth]{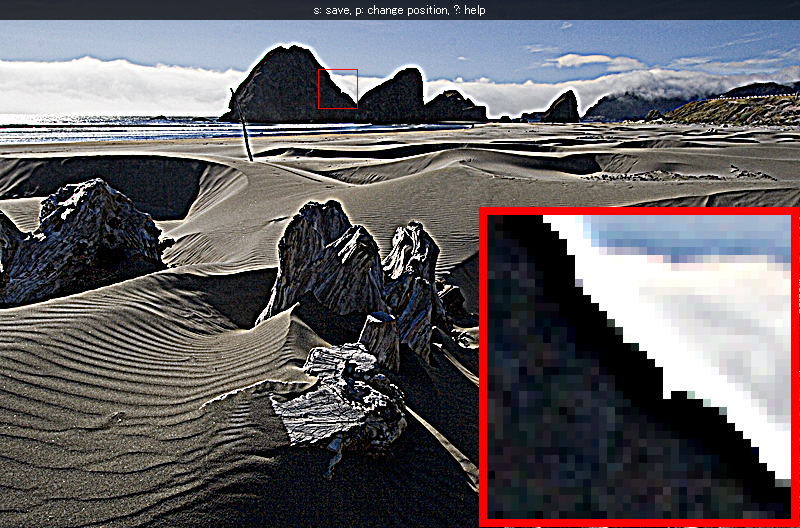}
  }
  \subfigure[MSBF]{
  \includegraphics[width=0.45\columnwidth]{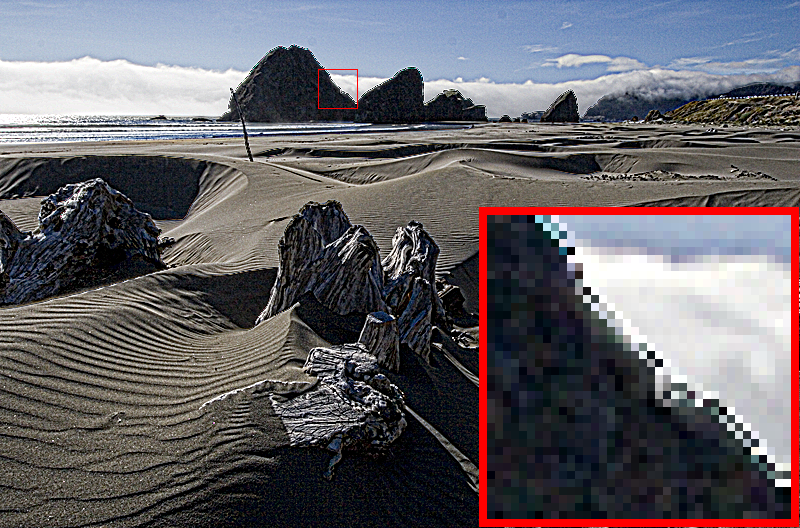}
  }
  \subfigure[EAW]{
  \includegraphics[width=0.45\columnwidth]{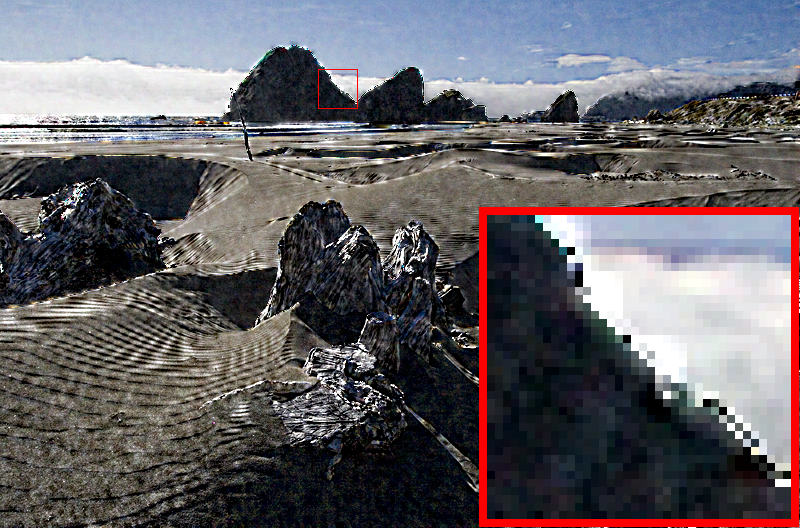}
  }
  \caption{Visual comparison with various method for Beach image (3-layers, $\sigma_r=30$, $m=7$, $K=10$).}
  \label{beach}
\end{figure}

Figures~\ref{flower} and \ref{beach} show the results of various enhancement methods.
For two-layers cases, Gaussian filtering makes large halos, but the other methods and proposed method suppress the halos.
The bilateral filtering generates edge-reversals, but the other suppress them.
Both domain transform and guided image filters can suppress halos and edge-reversals; however, both filters cannot enhance much.
The domain transform filter uses geodesic distance, which becomes large in complex texture regions; thus, the smoothing output in the textured region tends to keep the input image.
The guided image filtering uses linear fitting, which tends to ignore minor wave signals; thus, only significant edges in the image are enhanced.
For multi-layer cases, pyramid and MSBF enhancement are the multi-layers extensions of two-layers with Gaussian filtering and bilateral filtering; thus, the output and the problems are similar to the two-layer cases.
The EAW is a wavelet extension for MSBF; thus, the problems are also similar to the MSBF.
%, such as flower, stone building, motocross bikes, sailboat at anchor, shuttered windows, sailboats under spinnakers, mountain stream, lighthouse in Maine, P51 Mustang, Portland Head Light, and mountain chalet.

%\clearpage
%\IEEEtriggeratref{16}
\bibliographystyle{IEEE}
%\small
\bibliography{SPL}

\end{document}